\documentstyle[11pt,aaspp4,flushrt]{article}

\begin{document}

\title{A Tighter Test of the Local Lorentz Invariance of Gravity Using
PSR~J2317+1439}

\author{J.~F.~Bell} \affil{Mount Stromlo and Siding Spring
Observatories, Institute of Advanced Studies,\\ Australian National
University, Private Bag, Weston Creek, ACT 2611, Australia \\email:
bell@mso.anu.edu.au}

\author{F.~Camilo} \affil{Joseph Henry Laboratories and Physics
Department,\\ Princeton University, Princeton, NJ 08544 \and University
of Manchester, NRAL, Jodrell Bank, Macclesfield, Cheshire SK11~9DL, UK
\\ email:  fernando@pulsar.princeton.edu}

\author{T.~Damour} \affil{Institut des Hautes Etudes Scientifiques, F-91440
Bures-sur-Yvette, France \and D\'epartement d'Astrophysique Relativiste et de
Cosmologie, Observatoire de Paris,\\ Centre National de la Recherche
Scientifique, F-92195 Meudon, France\\ email:damour@ihes.fr}

\begin{abstract}
Gravity being a long-range force, one might {\it a priori} expect the
Universe's global matter distribution to select a preferred rest frame
for local gravitational physics. The phenomenology of preferred-frame
effects, in the strong-gravitational field context of binary pulsars,
is described by two parameters $\hat{\alpha}_1$ and $\hat{\alpha}_2$.
These parameters vanish identically in general relativity, and reduce,
in the weak-field limit, to the two parametrized post-Newtonian (PPN)
parameters ${\alpha}_1$ and ${\alpha}_2$.  We derive a limit of
$|\hat{\alpha}_1| < 1.7\times 10^{-4}$ (90\%~C.L.) using the very low
eccentricity binary pulsar PSR J2317+1439, improving by a factor of 3 on
previous limits.

\end{abstract}


\newpage

General relativity might not be the last word on gravity. On the one
hand, we do not know how to quantize it, and on the other hand, modern
unification theories suggest that the low-energy limit of unified
theories might contain, besides the tensor field postulated by
Einstein, other long-range fields participating in the gravitational
interaction between macroscopic bodies. The present paper follows and
improves upon previous results by showing how binary pulsar data can
set very tight limits on such possible long-range deviations from
general relativity.

In the weak-field context appropriate to describing the solar system the
parametrized post-Newtonian (PPN) formalism is used as a tool for analyzing
gravitation theory and experiment. It provides a set of parameters (the PPN
parameters) which take different values in different theories and can be
related to measurable quantities, forming a basis for comparison of theory
and experiment. See Will (1993)\nocite{wil93} for a summary of the PPN
formalism, PPN parameters, and the values they take in various theories of
gravity.  In theories where gravity is mediated in part by a long-range
vector field (or a second tensor field, besides the one postulated by
Einstein), one expects the Universe's global matter distribution to select a
preferred rest frame for local gravitational physics (violation of local
Lorentz invariance). In the post-Newtonian limit the phenomenology of
preferred-frame effects is describable by two PPN parameters ${\alpha}_1$
and ${\alpha}_2$. Gravitation theories, such as general relativity, which
respect local Lorentz invariance (absence of a preferred frame) have
parameters ${\alpha}_1 \equiv {\alpha}_2 \equiv 0$.  These two parameters
have been constrained using solar-system data $|\alpha_2| < 3.9\times
10^{-7}$, $\alpha_1 = (2.1 \pm 3.1)\times 10^{-4} $ (90\% C.L.;
\cite{nor87b}; \cite{hel84}; \cite{wil93}).

Binary pulsars take us beyond the weak-field context because the self
gravitational field of a neutron star is very strong. This strength is
conveniently measured by the ``compactness'' parameter $c_A = - 2 \partial
\ln m_A/ \partial \ln G$, with value $c_1 \simeq 0.3$ when $m_1 \simeq$
1.4\,M$_\odot$ (\cite{de92a}).  To discuss tests of gravity using binary pulsar
data, Damour and Taylor (1992)\nocite{dt92} introduced the parametrized
post-Keplerian (PPK) formalism in which a pulsar timing experiment is
parametrized by a set of phenomenological parameters which lump together all
possible strong-field effects.  The original PPK formalism (see also
\cite{ds91}) did not allow for preferred-frame effects.  Damour and
Esposito-Far\`ese (1992b)\nocite{de92b}, hereafter DE92b, extended the PPK
formalism by adding strong-field generalizations of the usual PPN
preferred-frame parameters, and showed how certain low-eccentricity binary
systems provided excellent testing grounds for possible violations of local
Lorentz invariance. Their parameters $\hat {\alpha}_1$, $\hat{\alpha}_2$,
are conceived as including the weak field contribution $\alpha_1$,
$\alpha_2$, plus possible strong-field modifications proportional to
successive powers of the compactnesses of the two stars: $\hat{\alpha}_1 =
\alpha_1 + \alpha_{1}^{'}(c_1 + c_2) + \cdot\cdot\cdot $, ; a similar
relation applies for $\alpha_{2}$.  In general relativity $\hat{\alpha}_1
\equiv 0 \equiv \hat{\alpha}_2$, while these parameters may be different
from zero in theories where gravity is partially mediated by a vector or a
second tensor field.  Binary pulsar data have already given a good limit on
$\hat{\alpha}_1$: $|\hat{\alpha}_1| < 5.0\times 10^{-4} $ (90\%~C.L.)
(DE92b).  Following the methodology of DE92b, we shall show here how the use
of the data on the very low-eccentricity binary pulsar J2317+1439 allows one
to significantly tighten this limit.\\

If $\hat{\alpha}_1 \neq 0$ a consequence is a constant forcing term in the
time evolution of the eccentricity vector of a binary stellar system. For a
very low eccentricity orbit, this tends to ``polarize'' the orbit, aligning
the eccentricity vector with the projection onto the orbital plane of the
absolute velocity of the system. Hence, the orbital parameters of very low
eccentricity binary pulsars may be used to set an upper bound on
$|\hat{\alpha}_1|$ (DE92b\nocite{de92b}), given by
\begin{equation}
\label{e:al1}
|\hat{\alpha}_1| < (10/ \pi) I_{i,\lambda} e/\hat{e}~~~~~(\rm 90\%~C.L.),
\end{equation}
where
\begin{equation}
\label{e:Iil}
I_{i,\lambda} = \int_{0}^{2 \pi} {(2 \pi )^{-1} d\Omega \over
\sqrt{1 - ( \cos i \cos \lambda + \sin i \sin \lambda \sin \Omega)^{2}}}
\end{equation}
and
\begin{equation}
\label{e:ecc}
\hat{e} = {1 \over 12} \left| { m_1 - m_2 \over m_1 + m_2 } \right|
{ \mid {\rm {{\bf w}_{psr}}} \mid \over (G (m_1 + m_2) 2 \pi/P_{b})^{1/3}}.
\end{equation}
In equations~\ref{e:al1}--\ref{e:ecc}, $e$ is the orbital eccentricity,
$\Omega$ is the longitude of the node of the binary orbit with respect to
the line of sight, $i$ is the orbital inclination, $\rm {{\bf w}_{psr}}$ is
the absolute velocity of the center of mass of the binary system, $\lambda$
is the angle between $\rm {{\bf w}_{psr}}$ and the line of sight, $m_1$ is
the pulsar mass, $m_2$ is the companion mass, $G$ is the gravitational
constant, and $P_b$ is the orbital period.

The very low eccentricity binary pulsar J2317+1439 (\cite{cnt93};
\cite{cnt95}) has a figure of merit for the determination of limits on
$\hat{\alpha}_1$ (DE92b\nocite{de92b}), 10 times better than PSR~B1855+09
which was used to obtain the previous best limit. Hence PSR~J2317+1439
should provide a substantially tighter limit on $|\hat{\alpha}_1|$. To
obtain the magnitude of the absolute velocity $| \rm {{\bf w}_{psr}} |$, the
cosmic microwave background is chosen as the preferred reference frame.  The
results from the Cosmic Background Explorer (\cite{sbk+91}; \cite{fcc+94})
for motion of the solar system with respect to the cosmic microwave
background give $|\rm {\bf w_{\odot}}| = 369$\,km\,s$^{-1}$ in the direction
$({\rm R.A., Dec}) = (11.3^{\rm h},-7\arcdeg)$.  This direction is almost
anti-parallel to the line of sight to PSR~J2317+1439 ($\lambda =
192\arcdeg$).  No estimate of the radial velocity of PSR~J2317+1439 is
available --- however Camilo, Nice, \& Taylor (1995),\nocite{cnt95} report a
velocity in the plane of the sky of $v_{\perp} = (70\pm
30)$\,km\,s$^{-1}$. Excluding, at the 90\% C.L., an unfavourable alignment
within $\theta_{min} = \arccos(0.9) = 25.8^{\circ}$ between $\rm {\bf
w_{\odot}}$ and minus the (solar barycentric) velocity of the pulsar, gives
a firm lower limit $| \rm {{\bf w}_{psr}} | > \sqrt{ ( \rm {{\bf w}_{\odot}}
- v_{\perp} \cot(\theta_{min}) )^{2} + v_{\perp}^{2} } \simeq 235
$\,km\,s$^{-1}$. The characteristic age of $23\times10^9$\,yr for
PSR~J2317+1439, and the cooling age for its stellar companion of greater
than $10^9$\,yr (\cite{cnt95}), ensure that it has been a long time since
mass transfer in the binary system ceased, satisfying the requirement that
systems used for this test be old (DE92b\nocite{de92b}).

Radio timing results give $P_b = 2.459$\,d and $e < 1.2\times 10^{-6}$
(90\%~C.L.) for PSR~J2317+1439 (\cite{cnt95}). Unfortunately no
measurements of the masses have been obtained to date.  However, since
all high-precision measurements of the mass of neutron stars are
consistent with $1.4$\,M$_\odot$, with very small scatter
(\cite{tamt93}), it is assumed that the mass of the neutron star in
this system is also 1.4\,M$_\odot$. The formation of binary systems
such as this one is sufficiently well understood to allow the
determination of $m_2$ from other orbital parameters (\cite{rpj+95};
\cite{jrl87}; \cite{sav87}). These methods yield $m_2 = (0.18\pm
0.02)$\,M$_\odot$, close to the minimum mass of 0.175\,M$_\odot$
derived from the pulsar mass function

\begin{equation}
\label{e:mfn}
f_1(m_1, m_2, i) = {(m_2 \sin i)^{3} \over (m_1 + m_2)^{2}} = { 4
\pi^{2} (a_{1} \sin i )^{3} \over G P_{b}^{2} } = 0.0022\,{\rm
M}_\odot,
\end{equation}
where $a_1$ is the semi-major axis of the pulsar's orbit. A 90\%~C.L.
upper limit on the mass is 0.45\,M$_\odot$ (\cite{pk94}).  Using these
masses, $i$ can be found from equation~\ref{e:mfn}.

Given the above parameters for PSR~J2317+1439,
equations~\ref{e:al1}--\ref{e:ecc} were evaluated, solving equation
\ref{e:Iil} numerically. This was done for several values of $m_2$ in the
range $0.175\,{\rm M}_\odot < m_2 < 0.45$\,M$_\odot$ to test the sensitivity
of the limit to changes in $m_2$. The resulting upper limits on
$|\hat{\alpha}_1|$ are in the range (1.2--$1.7)\times 10^{-4}$,
demonstrating that the limit is rather insensitive to the mass of the
companion and the inclination of the system.  The limit of $|\hat{\alpha}_1|
< 1.7\times 10^{-4}$ (90\% C.L.) is tighter by a factor of 3 than both the
previous best limit from pulsar systems of $|\hat{\alpha}_1| < 5.0 \times
10^{-4}$ (90\% C.L.; DE92b\nocite{de92b}), and the limit (probing only weak
field effects) from planetary data of $\alpha_1 = (2.1 \pm 3.1) \times
10^{-4}$ (90\% C.L.; \cite{hel84}), which allows an $\alpha_1$ as large as
$5.2 \times 10^{-4}$.

In principle the eccentricity of binary pulsars can be measured by
radio pulse timing techniques to about $\delta e \simeq \sigma/(a_1
\sin i)$, where $\sigma$ is the typical uncertainty in the measurement
of the absolute arrival time of a pulse.  For PSR~J2317+1439 $\sigma
\approx 1.5\,\mu$s, and $a_1 \sin i = 2.3$\,light-sec, consistent with
the present measurement precision of $e$.  Further improvement on
limits for $\hat{\alpha}_1$ from binary pulsars is likely to be slow,
since $\sigma$ for most pulsars is rather larger than for
PSR~J2317+1439, and the eccentricity for most is also much larger
(\cite{pk94}).  Nevertheless modest improvement can be expected as,
over time, the mass, velocity, and especially, eccentricity
measurements for PSR~J2317+1439 improve.  The limit we derived for
$\hat{\alpha}_1$ further constrains the strong-field regime of theories
of gravitation that deviate from general relativity in including extra
vector or tensor fields.

\acknowledgements

We thank Prasenjit Saha for helpful discussions.  JFB received support
from an Australian Postgraduate Research Award and the Australia
Telescope National Facility student program.  FC gratefully
acknowledges use of NSF grant AST~91-15103 at Princeton, and a
fellowship under the auspices of the European Commission while at
Jodrell Bank.

\clearpage

\bibliographystyle{apj1c}

\end{document}